# Estimating Hourly Neighborhood Population Using Mobile Phone Data in the United States

Huan Ning[1], Zhenlong Li[1] , Manzhu Yu[1], Shiyan Zhang[1] , Shan Qiao[2]

[1] Geoinformation and Big Data Research Laboratory, Department of Geography, Pennsylvania State University, University Park, PA, United States
[2] South Carolina SmartState Center for Healthcare Quality, University of South Carolina, Columbia, SC, United States

**Abstract**: Traditional population estimation techniques often fail to capture the dynamic fluctuations inherent in urban and rural population movements. Recognizing the need for a high spatiotemporal dynamic population dataset, we propose a method using smartphone-based human mobility data to reconstruct the hourly population for each neighborhood across the US. We quantify population fluctuations on an hourly, diurnal, daily, and seasonal basis, and compare these with static population data to highlight the limitations of traditional models in capturing temporal dynamics. This study is one of the first hourly population products at a large geographic extent (US), contributing to various studies that involve dynamic populations with high spatiotemporal resolution, such as air pollution exposure analysis and emergency response.

## 1 Introduction

Population data is fundamental for various studies involving human society, such as urban planning, public health, social sciences, resource distribution, and emergency management. Administrators invest vast amounts of effort in obtaining the population data in a timely manner to understand how many people live and work in jurisdictions, as well as their demographic characteristics (Baffour, King, and Valente 2013). Generally, knowing the population distribution and its trends could help society better allocate infrastructures and other resources to meet its needs and provide information to predict further patterns.

Traditional data collection methods, such as censuses conducted every decade in the United States (US) and China, offer detailed insights but often suffer from significant time lag, with data aggregation and processing taking up to 2 or 3 years (Bureau 2022). To mitigate this lag, intermediate surveys, such as the American Community Survey 1-year and 5-year estimates (ACS) at the Census block group (CBG) level, are utilized. These products contain population size, distribution, demographic characteristics, and socio-economic information; they are widely used in society management and science research (Hattis 2020). The spatial resolution of Census data can be very detailed since it surveys every household, but such data will not be released due to privacy concerns. The commonly used fine-grained product level is at the CBG level with each unit containing 1000 – 3000 residents, aggregated from minimum areal unit of blocks.

Contemporary societies are dynamic, where population distribution changes geospatially and temporally. For a specific region, the number of residents may increase or decrease due to economic development, environmental changes, newly built housing units, or migration. These changes can be on the metropolitan or neighborhood scale, in decades or years. The most dynamic population change is the commuting patterns, which refers to individuals traveling from home to the workplace. For example, the residents living in "sleeping cities" or "bedroom communities" will travel to employment centers or central business districts during the daytime on weekdays. Such highly dynamic daily mobility patterns are common in human society, especially in urban areas. Many city facilities are designed to accommodate these patterns. For instance, commuting patterns shape the transportation systems, including subways, buses, trains, and private vehicles. The commuter population size is one critical factor impacting the transportation infrastructure. Similarly, urban facilities, such as public parks, grocery stores, and healthcare facilities, are profoundly influenced by the daily and hourly changes in the nearly dynamic populations. The traditional static estimation of the resident population lacks the necessary information to help plan those city facilities.

The surge in population during events also needs to be estimated for safety concerns. Such estimation could be life-critical. For example, on the evening of October 29, 2022, a tragic crowd crush occurred during Halloween festivities in Itaewon, South Korea (Mao 2023; Sharma et al. 2023), killing 150 people. On December 31, 2014, in Shanghai, China, 36 people died, and 49 were injured in the New Year's Eve stampede (Zhou, Pei, and Wu 2018). Religious gatherings are also events that may cause the population to increase dramatically (Gayathri, Aparna, and Verma 2017). Dynamic population estimation can provide data support for authorities to control the crowd, avoiding crushes. For hazard exposure estimation such as air population, using dynamic hourly population in line with the real-time observation data may provide a more accurate estimation than resident population (Gariazzo, Pelliccioni, and Bolignano 2016; Picornell et al. 2019; Yu et al. 2023; 2024).

The dynamic population is caused by human mobility, such as commuters (workers or students), tourists, hospital patients, shoppers, and club-goers. Moss and Qing (2012) synthesized multiple data sources to estimate the dynamics of the Manhattan population. Their study showed that there were 3.94 million people on weekdays in 2010, while the Census reported 1.62 million residents and 3.07 daytime population. The primary data sources contain the commuter and resident data from the Census, ACS, college students, hospital patients, and statistics reports on counts of visitors. This early study shows that there is a significant gap between the estimated dynamic population and the resident population.

Studies have been using mobile phone data to capture the population dynamics. One widely used mobile phone mobility data is the call detail record (CDR), which records the time the mobile phone communicates to the tower when phone users make calls or texts. Such data is often viewed as population proxies due to the high prevalence of mobile phones. For instance, the prevalence of smartphone use in the US will be 97% in 2022. Mobile phone-derived population can be an alternative for areas without reliable census data, such as low-income areas (Deville et al. 2014).

Deville et al. (2014) used 5 – 10 months' 1 billion mobile phone call records from Portugal and France to estimate the real-time population. The dataset contains the served mobile phone count of phone towers and user pairs when they made a phone call or text. The authors correlated the population density and the mobile phone density surrounding the phone tower. They then used this correlation to convert the known mobile phone density to population density. Similarly, Liu et al. (2018) adopted one-day mobile phone call records in Beijing, China, for hourly population density estimation. One additional step in that study was reconstructing the individual trajectories so the authors could generate a relatively accurate phone device density at hourly intervals. Since such studies are based on mobile phone call records or network connection records, they are subject to the carrier's tower coverage.

Bergroth et al. (2022) consider more factors, mostly the building floor areas, for dynamic population estimation. They applied the building floor count to estimate the accommodation areas rather than the land area, then estimated the weekdays, Saturdays, and Sundays 24-hour dynamic population via such dasymetric interpolation. Wei, Su, and Liu (2023) noted the different human activities in land cover types in their dynamic population activity analysis. LandScan provides daytime and nighttime population only based on various data sources, such as land cover (Weber et al. 2022).

Although many endeavors have been made to increase the geospatial resolution of populations (WorldPop 2024; Huang et al. 2021; Weber et al. 2022), relatively few attempts have been made to determine temporal population dynamics. The former usually adopt the land cover and land use data, such as building footprints, to allocate the population into smaller units. The dynamic population estimation is limited by data availability. Transportation and mobile phone data are two major data sources, however, they often vary among study areas due to various data providers and limitations. While existing studies used geotagged tweets to estimate daily population at the county level (Martin et al., 2021; Huang et al., 2022), universal methods for large-area estimation of population dynamics with a high spatial (e.g., CBG level) and temporal (e.g., hourly) resolution are rare in the literature.

To bridge this gap, we propose a method using smartphone-based human mobility data (SafeGraph Patterns; now Advan Patterns) to reconstruct the hourly population for each neighborhood (Census block group) across the US. This study produced one of the first hourly population datasets for the entire US, serving various studies that involve dynamic populations at fine-grained spatiotemporal scales. Four levels of dynamic population patterns are observed in the results, including hourly, diurnal, daily, and seasonal, reflecting various local events, such as campus crowds, festival parades, and tourist flows. These patterns are valuable supplements for static populations such as LandScan and ACS annual estimates.

## 2 Results

In this section, we present four observed dynamic population patterns: hourly, diurnal, daily, and seasonal. Census block groups' populations were aggregated into towns or cities according to the 2023 US Census places boundary. All block groups having more than 50% of the area located in the place boundary will be aggregated. We plotted and mapped the aggregated population and visually inspected them for patterns.

### 2.1 Hourly patterns

The hourly population clearly shows the peak hours and off-peak hours. In the morning, commuters run to the workplace, such as central business districts, and return to residential communities when off work. Figure 1 displays the State College borough, Pennsylvania, a university town in which the Pennsylvania State University is located. The upper highlighted areas among the 27 Census tracts is the campus. The hourly population plot of this university town reveals the students gathering during the daytime. There are about 49,000 enrolled students and 7,000 full-time faculty and staff in this campus. Figure 1 displays a reasonable estimation of the daytime campus population.

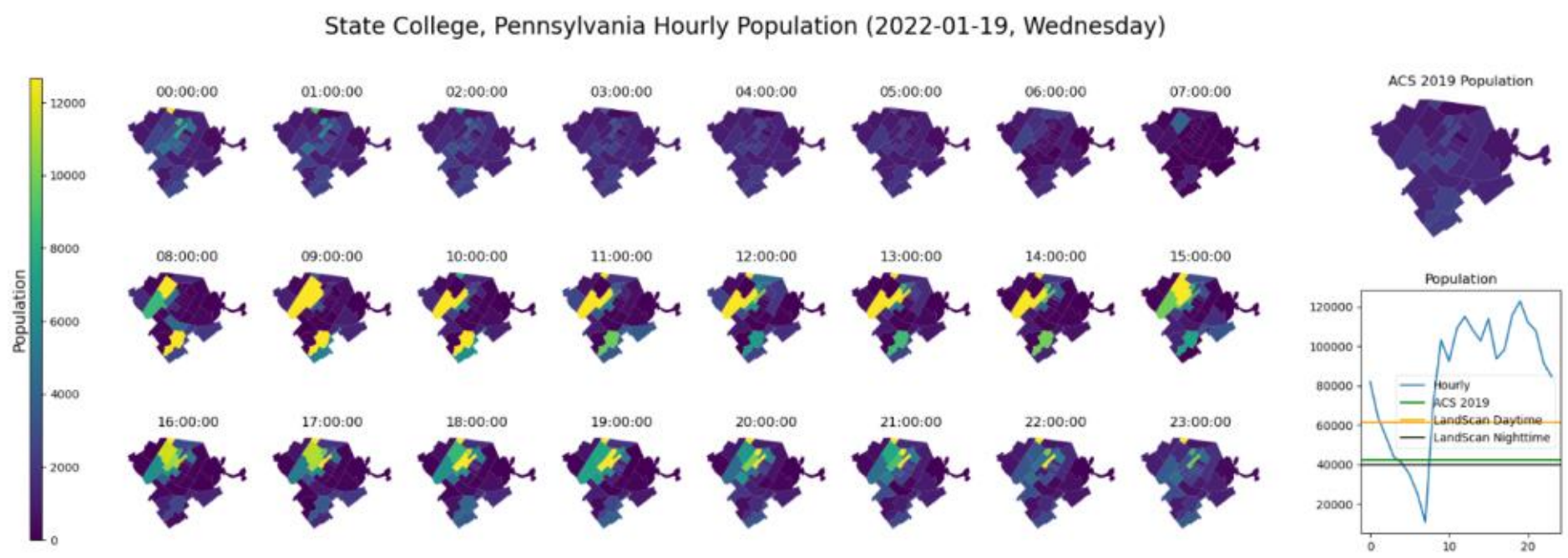

Figure 1 Hourly population of State College borough, PA, on Wednesday, January 19, 2022. Each polygon is a Census block group. This borough is a university town for Pennsylvania State University. The upper highlighted tracts are the campus.

## 2.2 Diurnal patterns

The residential neighborhoods and business districts might appear to have diurnal patterns that are opposite to those of the population. Our results provide a clear view of such patterns, showed in Figure 2. Tracy city, California, a bedroom community for the Bay Area, where commuters leave home in the morning and back home at night (Figure 2(a)). Both dynamic and LandScan population reflects this pattern: the daytime population is less than the nighttime, and the nighttime population is similar to the ACS population. Understanding such spatio-temporal patterns helps urban planners and city managers better build related infrastructures, such as highway systems and healthcare facilities. Commerce city, California, is an opposite example of Tracy city; it is an industrial area, and its daytime population is higher than the daytime population (Figure 2(b)). The landscape of these two cities can be found in the right column of Figure 2; Tracy city is a large and topical residential community, while Commerce city is an industrial area full of factory workshops.

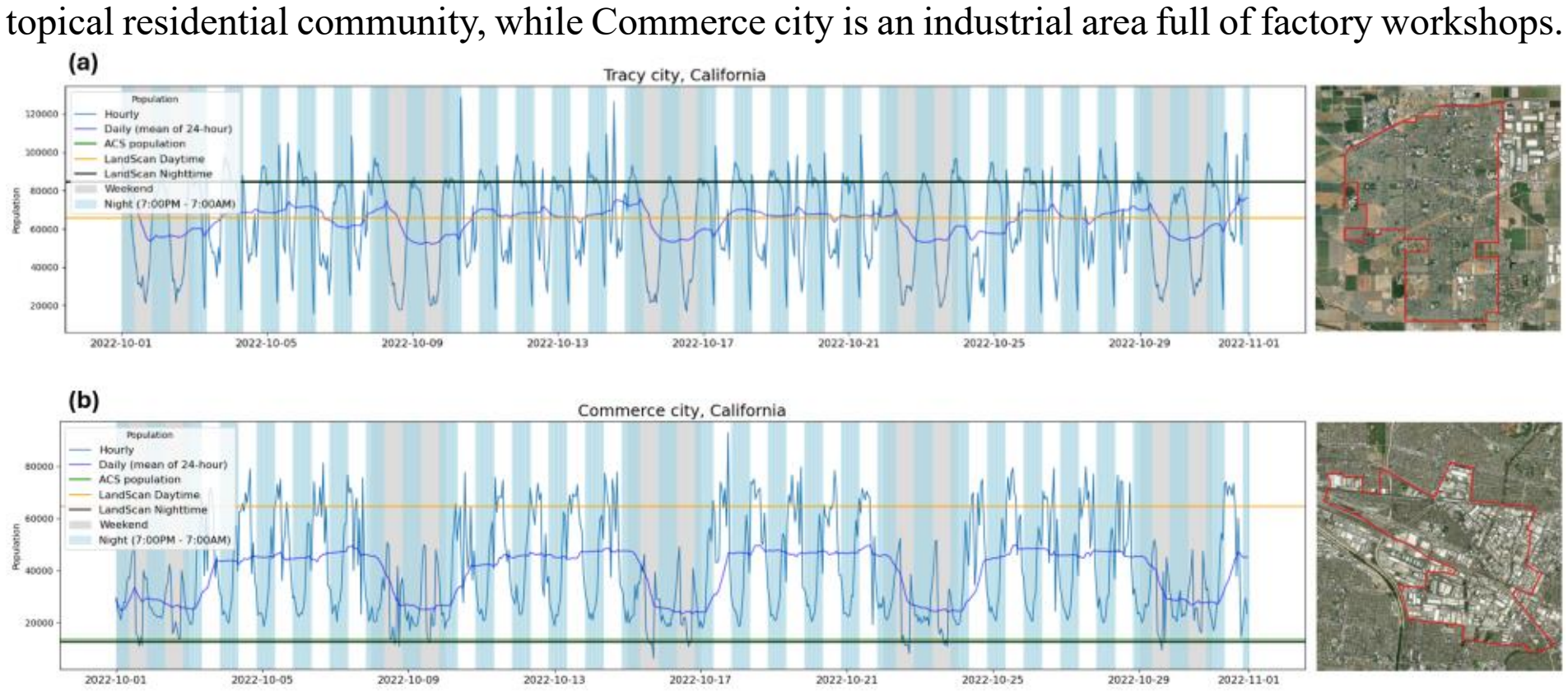

Figure 2 (a): The diurnal population trend in Tracy city, California. This city is a bedroom community for Bay Area commuters. The hourly population plot shows a daily pattern: the population dropped during the daytime (7:00AM-7:00PM), and increased during nighttime. Note that the horizontal lines of the ACS population and LandScan Nighttime are overlapping. (b):The diurnal population trend in Commerce city, California. This city is an industrial area. The hourly population plot shows a daily pattern: the population increased during the daytime and dropped during nighttime.

### 2.3 Daily patterns

We used a moving window of 24 hours to average the hourly population as the daily population. Many places show a weekly periodic pattern: people go to workplaces on weekdays and stay at home at weekends. Figure 3 presents a topical city, Havelock, North Carolina, which had such a stable daily pattern. A large Marine Corps air station, Marine Corps Air Station Cherry Point, is located in this city. The relatively stationary military routine may cause such an accurate population daily pattern. Figure 3(a) shows the differences between weekdays and weekends in the year 2022. The population of October 2022 is zoomed in in Figure 3(b); the weekday daily population is close to the LandScan daytime population at about 15,000, while the weekend population is similar to the LandScan nighttime population at about 9,000. However, the peak hourly population in the daytime achieved up to 35,000. Notably, the daily population curve reflects the Labor Day holiday on September 5; on that Monday, the daily population is similar to weekends rather than weekdays, while deceased before and after that long weekend.

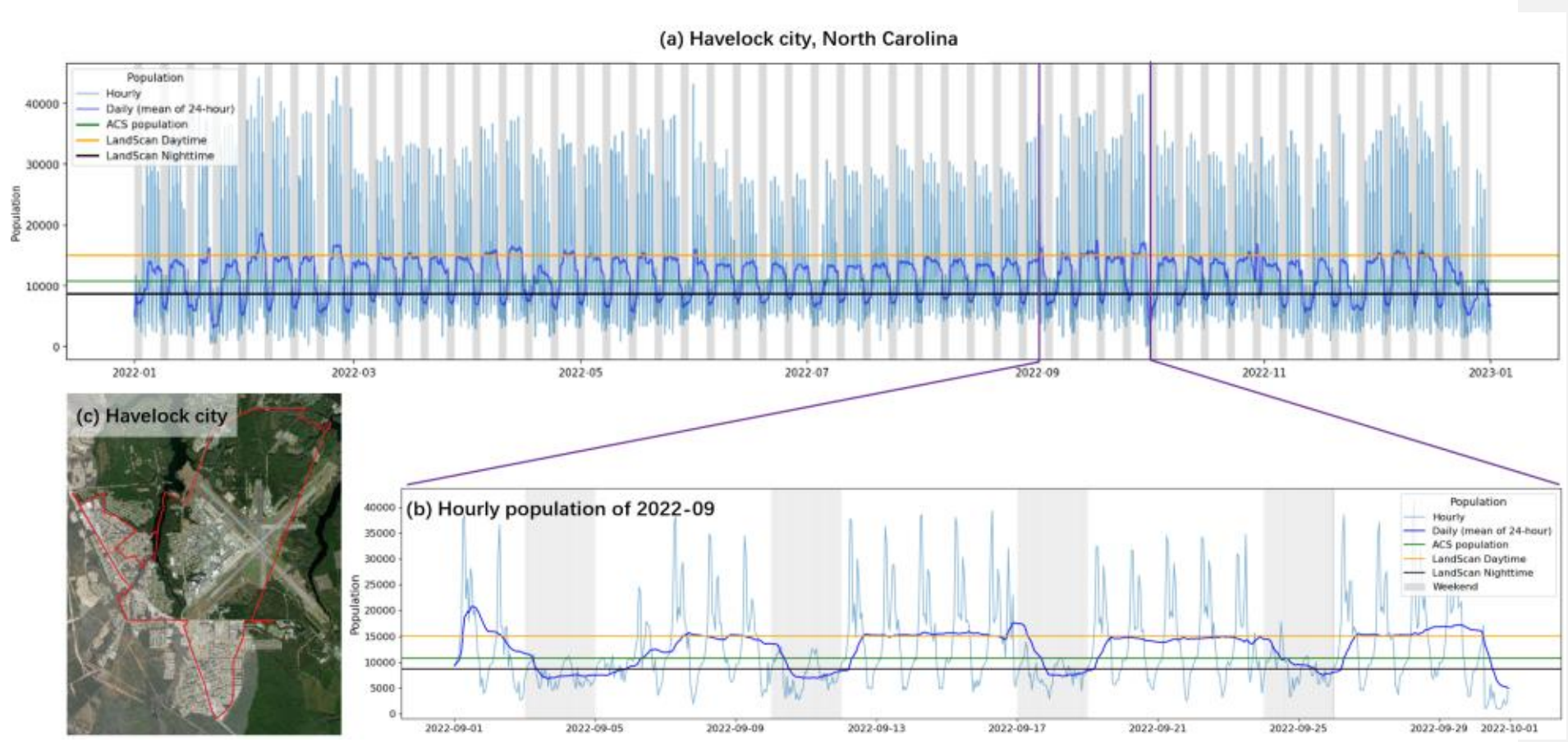

Figure 3 (a): The daily population dynamic pattern of Havelock city, North Carolina. (b): The weekdays population is much higher than the weekends and holidays, such as Labor Day (September 5). (c): the landscape of Havelock city.

Daily population data is critical for emergency response. For example, on September 10, 2022, Hurricane Ian landed in South Carolina, 50 km away from the famous coastal tourist city of Myrtle Beach. How many people would be impacted? Figure 4 may provide the answer: most tourists and residents have fled out of that attraction. Such information is valuable for disaster management in similar circumstances.

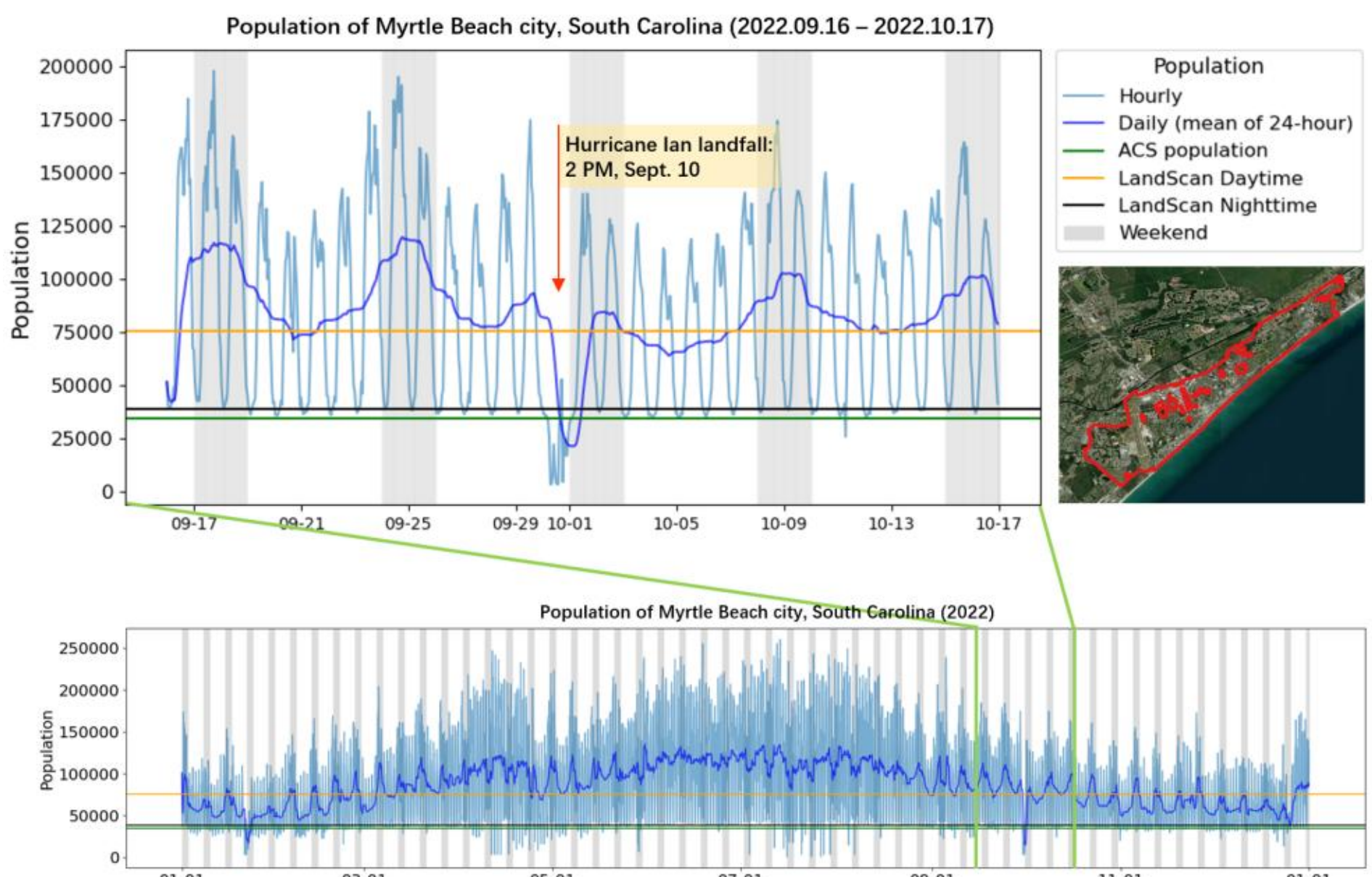

Figure 4 Visitors go to this city for the beach all year round, especially in the summer. However, they evacuated out of Myrtle Beach city due to the landfall of Hurricane Ian on September 10. The sudden population drop in the research results provides critical data for emergency response.

Some local festivals may attract remarkable visitors, and these days can be highlighted in our study. Figure 5 shows two local events in Benson town, North Carolina. During the traditional Mule Days (September 22 – 24), Benson town's population peaked yearly on Saturday, September 24. About 20,000 – 30,000 visitors came to the festival (Johnston County Visitors Bureau 2024), while the peak population of our result is about 25,000, close to the reported visitor count. The Christmas parade also drew a similar amount of visitors on December 2. Such dramatic population changes can be reflected in the daily population. On the contrary, ACS and LandScan data record less than 1000 population for the entire year, failing to capture the surging population during festival events.

Similarly, about 9,000 people visited the 45th Annual World Catfish Festival on Saturday, April 2, 2022, in Belzoni city, Mississippi (World Catfish Festival 2022). Our results show the peak population is about 25,000. Although the numbers are inconsistent, the dynamic population captured the population spike.

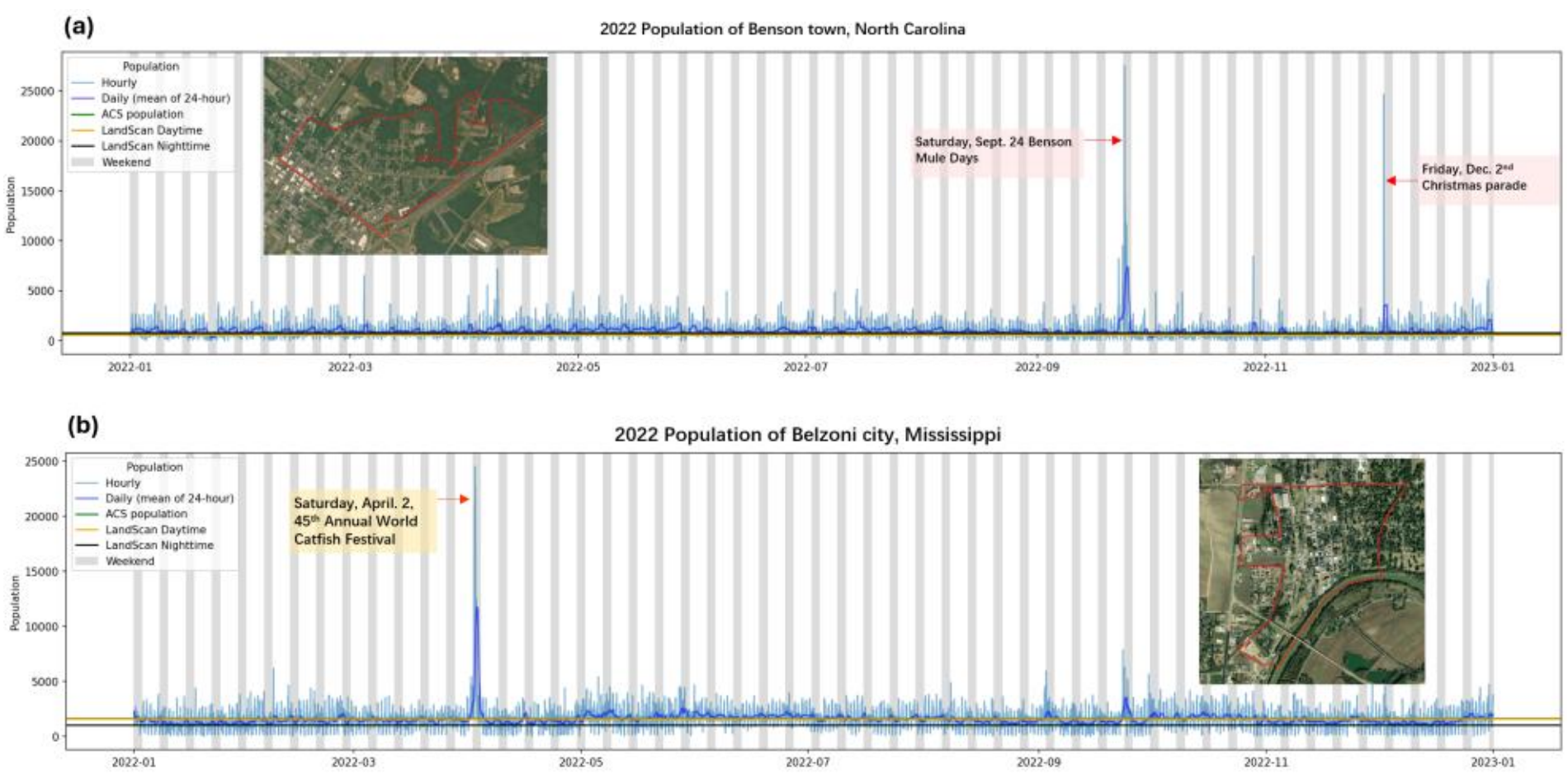

Figure 5  (a):2022 population of Benson town, North Carolina. (b): 2022 population of Belzoni city, Mississippi

## 2.4 Seasonal patterns

Seasonal population movements refer to the regular, predictable shifts in human populations based on seasonal factors such as climate, holidays, agricultural cycles, and cultural events. Understanding such population dynamics benefits various fields, such as infectious disease spreading and urban planning (Wesolowski et al. 2017; Woods et al. 2022; Buckee, Tatem, and Metcalf 2017). Our experimental results clearly showed the population changes among seasonals, including school semesters, local holidays, and tourist seasons.

Figure 6 (a) presents the university gatherings and dimissings via the estimated dynamic population of Pennsylvania State University. The changing population curve accurately matches the university academic calendar (Penn State 2022): during the spring, summer, fall, and winter breaks, the town population drops dramatically. The ACS 2019 and LandScan nighttime population meet the summer break population but fail to represent other periods.

The study results captured the seasonal tourist flows to attractions. For example, tourists crowd into a small coastal town, Holden Beach town, North Carolina, during the hot summer but avoid it in the windy and cold winter. Figure 6 (b) clearly presents the tourist flows, including peaks during school breaks and national holidays. For example, the long weekend of Memorial Day (Monday, May 30) and Labor Day (Monday, September 5) resulted in more visitors. Both ACS and LandScan data do not consider the tourist flows, and they show a small population of about 1,000 here all year round.

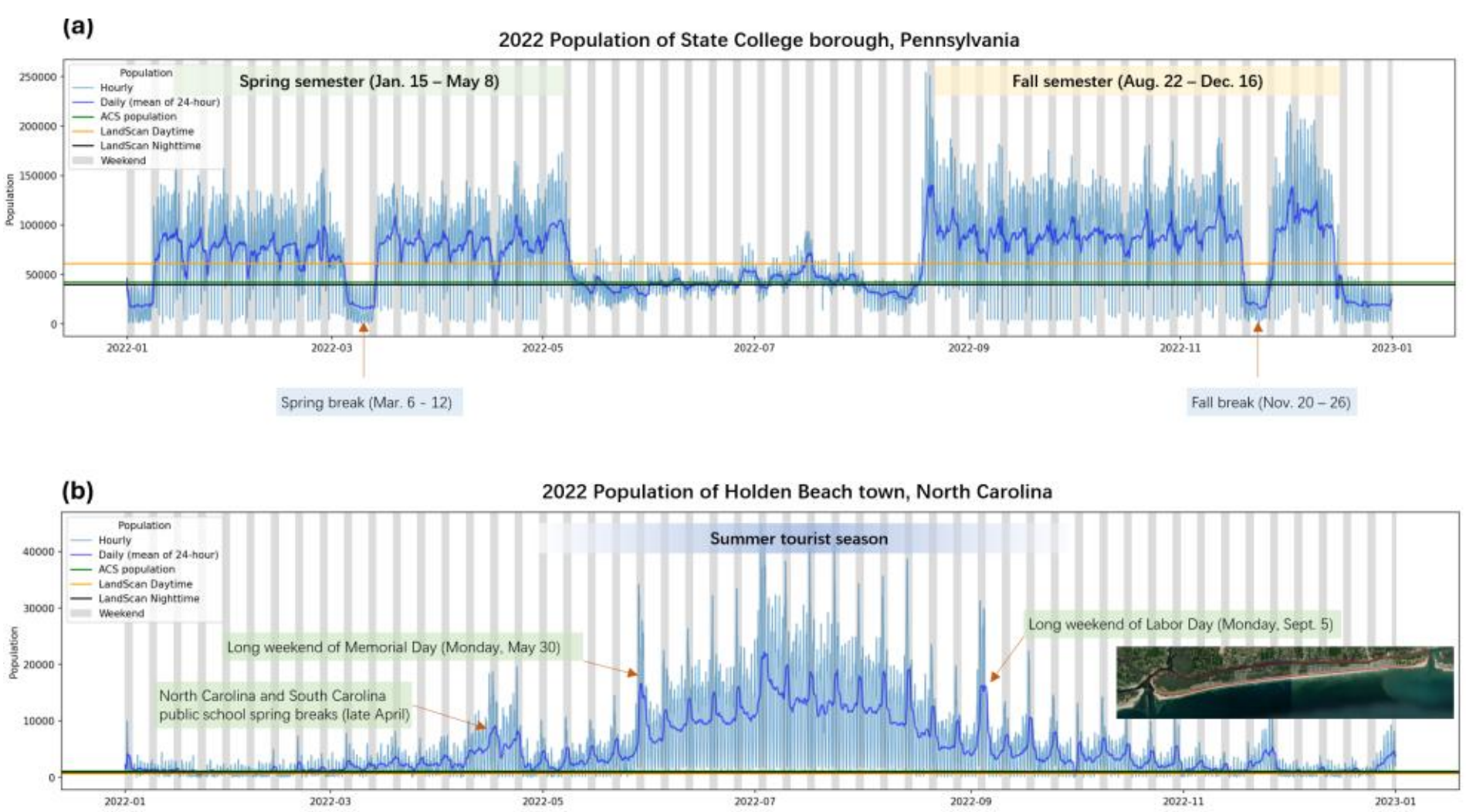

Figure 6 (a) Seasonal population changes in State College, PA. This borough is a university town for Pennsylvania State University. The dynamic population clearly shows the university gatherings and dismissings. (b): Seasonal tourist flows of Holden Beach town, North Carolina. Tourists crowd into this small coastal town during the hot summer but avoid it in the windy and cold winter. Our study results clearly present the tourist flow, including peaks during school breaks and national holidays. For example, the long weekend of Memorial Day (Monday, May 30) and Labor Day (Monday, September 5) resulted in more visitors. Both ACS and LandScan data do not consider the tourist flows, and they show a small population of about 1,000 here all year round.

## 3 Discussion

The research results are promising since the population is reasonable compared with other sources. Similar to other population products (Moss and Qing 2012; Weber et al. 2022), this study is based on the Census block group data, meaning that the derived population would not depart the Census data significantly. Due to a lack of sufficient calibration and validation, our results may contain various uncertainties from data collection, modeling, post-processing, and evaluation. Although this is a quantitative study, we suggest that the readers interpret the numbers in a reasonable manner and do not expect these hourly populations to be highly accurate. Uncertainty also comes from the aggregating boundaries and time periods. We found that the dynamic population patterns with an aggregated population of about dozens of thousands for a place of medium area tend to show clearer patterns. A large region may cover most of the destination and origin Census block groups, so the aggregation population appears smooth without observable patterns. A small region with one or a few Census block groups might be noisy. Due to the high spatiotemporal resolution of our dynamic population products, we recommend against applying these data indiscriminately across large metropolitan areas where the aggregation might mask important local variations. Instead, larger areas can be segmented into smaller, manageable sections to analyze and interpret the data more effectively. Also, it is essential to select appropriate temporal resolutions - hourly, diurnal, daily, or seasonal – that align with the specific phenomena of interest.

Our experiment showed that the hourly population can be derived from current publicly accessible datasets of mobile phone data, and it can be a standard public product. Such products can provide high spatiotemporal resolution, capturing the dynamic population patterns at hourly, diurnal, daily, and seasonal levels. The ability to monitor population movements on an hourly basis has wide-ranging applications in various domains, such as infrastructure development and emergency response. For example, the successive hourly population change in different neighborhoods is a critical variable for transportation simulation. Another application of the hourly population is to facilitate hazard assessment. Some pioneer studies (Gariazzo, Pelliccioni, and Bolignano 2016; Picornell et al. 2019; Yu, Masrur, and Blaszczak-Boxe 2023) have demonstrated that the dynamic population helps to obtain accurate hazard exposure such as wildfire smoke (Yu et al. 2024) and hurricane. The diurnal patterns shown in Tracy ciy and Commerce city can provide data support for transportation and other infrastructure improvements (Section 2.2). Crow safety monitoring, such as the Benson town Christmas parade and Belzoni town World Catfish Festival, is another application if the dynamic population can be obtained in real-time, which is possible (Section 2.3).. The seasonal patterns of tourist flows can help visitors and tourism operators better plan their trips and business (Section 2.4).

Some limitations are worthy of being noted. The first concern is the quality of phone-based datasets. SafeGraph Patterns datasets are relatively new (the earliest dataset is 2018), and they are still being developed. It is expected because the GPS-enable smartphone predominance were started merely a decade ago, while the raised location privacy issues, changing user policies, and regulation impact the data collection. In addition, large-scale smartphone-based data are usually compiled from various partners and software developer toolkits (SDKs); these data sources are unstable and rapidly changing, so data quality control is difficult and lacks transparency. For example, in our experiment, the device sampling rate is critical, but the 2023 data changed dramatically from the historical data. Although the data provider restated the data, researchers have no accessible resources to evaluate the device sampling rate. The second concern is the drawbacks of the modeling method. We adopted several assumptions that might not accurately reflect real-world conditions. For example, the hourly origin CBG distribution is considered the same as the month origin CBG distribution - a simplification that does not account for daily variations in human mobility, such as differences between weekdays and weekends. Additionally, the quantitative relationship between the tracked devices, device stops, and visitors is challenging to model. If the data provider can release detailed datasets, including the device trips with destination and origin CBGs, the data limitation will be largely mitigated.

The core idea of using mobile phone data to estimate dynamic population is to convert the observed phone count to headcount. Assessing such conversion requires dynamic population datasets or proxies; however, such data is rare. Researchers can create these datasets to evaluate the dynamic population products and establish benchmarks, such as using high school student counts or other known foot traffic data to calibrate the mobility data. Besides tackling data limitations, another future research avenue can be improving modeling methods. One research avenue is human mobility simulation (Replica 2024). Each resident can be simulated using an agent that follows the observed data, such as device stops and POI visitation. High-fedility simulations can provide rich data products for various human mobility research, such as traffic optimization and epidemiology control.

## 4 Conclusion

In this study, we investigated the possibility of deriving the hourly population at the neighborhood level for the US based on mobile phone data. The results provide a reasonable approximation compared to traditional static population data sources like LandScan. The relative differences observed - 67.8% during daytime and 42.7% at nighttime compared to LandScan grided population products at the Census block group level - highlight the dynamic nature of population movements that traditional censuses and static models may overlook. The detailed examination of dynamic population patterns across hourly, diurnal, daily, and seasonal scales show consistent trends with real-world events and demographic shifts. The increased population during university sessions or tourist peaks in coastal areas aligns closely with expected human activity patterns. Although the accuracy of these results is subject to data limitations, appropriately interpreting such patterns is valuable for various applications, such as infrastructure planning and emergency response.

## 5 Methodology

### 5.1 Study area and data

We chose the United States as the study area, covering 331 million people. This is one of the largest populations among similar studies. Many commercial companies provide data products reflecting human mobility patterns, such as Cuebiq (cuebiq.com), Replica (replicahq.com), and Pass_by (passby.com). We adopted the widely used SafeGraph Patterns data to estimate the hourly population.

The Advan Neighborhoods Patterns (Advan Research 2022) record the tracked mobile devices' stops in CBGs. SafeGraph and its partners track mobile devices (e.g., smartphones) anonymously in their device panel by recording the devices' stops inside CBGs. A stop refers to a device that stays in a CBG for more than 1 minute. Neighborhood Patterns contains the raw stops in CBGs in a calendar month, the hourly stops in CBGs, and, for each CBG, the home CBGs of the visited devices and the associated counts. SafeGraph also provides the panel device counts for each CBG to compute the device sampling rate. The data period is 2022.

The other major data source used in this study is ACS 5-year estimates 2019. The ACS data year is different from the mobility data; there are newer ACS datasets, but they are based on 2020 Census CBG boundaries, not aligning with the Neighborhood Patterns based on 2010 Census boundaries. We did not find a reliable and relatively simple method to convert the mobility data between different CBG boundaries, so we chose to use ACS 2019. The latest ACS 5-year estimates are 2021 during this research, and the entire population increased by about 1.8% (325 vs. 331 million), so ACS 2019 data will not affect the result and conclusion to a large degree.

### 5.2 Hourly population estimation

To estimate the dynamic population of a Census Block Group (CBG) on an hourly basis, we developed a model that accounts for both incoming and outgoing individuals. Using a straightforward method, we calculated the hourly CBG population based on the ACS population. Specifically, for each origin CBG, denoted as $c$, its hourly population, $p_c^t$, can be computed by: 1) subtracting the resident count left home (outbound), and 2) adding the visitor count (inbound) at that hour to its population, see Equation (1):

$$p_c^t = N_c - Out_c^t + In_c^t \qquad (1)$$

where,

$t$: hour $t$ in the month, $\tau$ in total. For example, if a month has 30 days, $\tau$ is 720 (24 hours/day * 30 days).

$c$: Census CBG, $n$ in total.

$p_c^t$: count of people in Census $c$ at hour $t$.

$N_c$: resident population (ACS) of CBG $c$.

$Out_c^t$: outbound, count of people who left their home CBG $c$ at hour $t$.

$In_c^t$: inbound, count of people who visit CBG $c$ at hour $t$.

In Equation (1), the hourly population ($p_c^t$) is we are looking for, and the CBG population ($N_c$) can be checked out from ACS data; thus, we need to find inbound ($In_c^t$) and the outbound ($Out_c^t$).

### 5.2.1 Estimate hourly inbound ($In_c^t$)

SafeGraph Neighborhood Patterns dataset provides the hourly stop counts of a CBG. A stop means that a tracked device stays more than 1 minute in the CBG. Then, one crucial step is converting the device count to the actual inbound, i.e., visitor count. Patterns datasets contain the device sampling rate for each CBG. Thus, if we know the home (i.e., origin) CBG distribution of these stopped devices in the destination CBGs, we can compute the actual visitors according to the CBG population, device sampling rate, and stopped device count. The Neighborhood Patterns dataset provides the monthly stopped device count and their origin CBGs; we assume that the hourly home CBG distribution is the same as the monthly distribution, so that the hourly inbound $In_c^t$ can be obtained by Equation (2):

$$In_c^t = \sum_{j=1}^{n} C_c^t r_j^t P_j^m\, k \qquad (2)$$

where,

$C_c^t$: stopped device count in CBG $c$ at hour $t$, provided in the Neighborhood Patterns dataset. Note that the dataset provides hourly stop count and median device dwelling time, meaning that if a device stayed for more than 1 hour in a place, the recorded stop count may be only once, so we added extra stops for those CBGs having device dwelling time more than 1 hour. For example, if a CBG's device dwelling time is 1.8 hours and there were one stop at hour *t*, we would add an extra stop at the following hour $t + 1$. We assumed that a stop indicated one device only, so the hourly stop count from the Neighborhood Patterns dataset represents the newly arrived device count at that hour. However, we found that the data mostly cannot record the long dwelling time correctly; the possible reason is that the mobile device may turn off the positioning function, and then SafeGraph’s partners have difficulty collecting the actual dwelling time. The SafeGraph documentation also mentioned that they can only distinguish a limited number of workers (SafeGraph 2023). Thus, the estimated population after the morning peak hour may be inaccurate.

$r_j^t$: the ratio of stopped devices coming from each CBG $j$ at hour $t$. Although $r_j^t$ is not provided in the Neighborhood Patterns dataset, we assume that $r_j^t$ equals the monthly ratio $r_j^m$, where $r_j^m = d_j^m / R_c^m$, $d_j^m$ is the stopped devices originated from CBG $j$ in month $m$, $R_c^m$ is the total stopped devices to CBG $c$ in month $m$. $d_j^m$ and $R_c^m$ were provided in Neighborhood Patterns.

$C_c^t r_j^t$: the stopped device count comes from CBG $j$. The stopped devices $C_c^t$ is a combination of devices originating from different CBGs; one device may represent different amounts of actual people since Neighborhood Patterns have different device sampling rates among CBGs. For example, it might track 10% of devices in a CBG and track 2% in another CBG; Figure 7 is a diagram showing the components of $C_c^t$ and $In_c^t$. Thus, we need to decompose $C_c^t$ into their origin CBGs (i.e., CBG $j$), then use their associated device sampling rate to estimate the actual inbound $In_c^t$, i.e., visitor count.

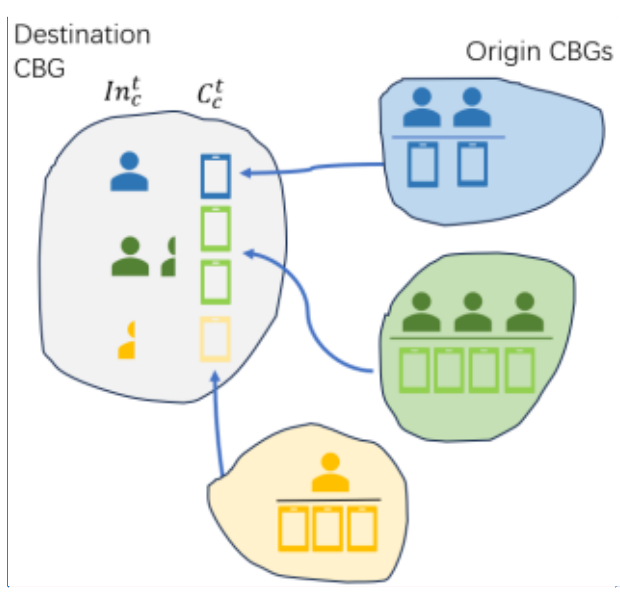

Figure 7 The stopped devices' destination and origin CBGs. At hour $t$, the stoppted devices ($C_c^t$) in CBG $c$ consists of devices origined from different CBGs. Each device may represent different amounts of outbounds since the device sampling rates vary among CBGs. For instance, in this figure, a blue device represents one person, a green device represents a 3/4 person, and a yellow one represents a 1/3 person. Note that these numbers are arbitrary for easy understanding; the actual number of people per device is about 50 – 5, according to the device sampling rate (2%-20%).

$P_j^m$: the number of people indicated by a tracked device in month $m$. $P_j^m = N_j / D_j^m$, where $D_j^m$ refers to the tracked device count in CBG $j$; it was provided in the Neighborhood Patterns. $P_j^m$ is also the reciprocal of device sampling rate of CBG $j$.

$k$ : the adjusting coefficient for unmeasured uncertainty. The collected mobile phone data may not fully reflect human mobility. For example, not all panel devices were tracked throughout the entire month, and not all the stops would be recorded. Thus, we used this coefficient to model various uncertainties in total. It is difficult to determine a suitable value for $k$, especially for different CBGs and hours across a month. We used the trial-and-error method to test various values and then compared the results with previous research in Manhattan, New York (Moss and Qing 2012; Weber et al. 2022). We could not find similar calibrating or control areas in the literature, so we cannot assess the $k$ or tailor it for difference regions and periods. The two previous studies (Moss and Qing 2012; Weber et al. 2022) contain the estimated population of weekday daytime, weeknight, weekend daytime, annual daytime, and annual nighttime. Using a trial-and-error method, we set $k = 4$ (see Section 5.4). This is a rough and universal estimation for the entire

**Commented [ZL1]:** The human icon in the figure is covered.

**Commented [HN2R1]:** No.

nation and hours in 2022; it may bring extra uncertainty, but we do not have a better calibration yet.

### 5.2.2 Estimate hourly CBG outbound ($Out_c^t$)

The Neighborhood Patters dataset does not contain the CBG outbound $Out_c^t$. It is tricky to estimate it. Essentially, the mobility flow can be formatted as a spatiotemporal mobility cube, or a mobility matrix of three dimensions, $M$, while $M \in \mathbb{R}^{\tau \times n \times n}$; the three dimensions are $Time$, $Destination$, and $Origin$. Each element of this matrix is the visitor count that comes from $origin$ CBG ($\mathrm{CBG}_j$) to the $Destination$ ($\mathrm{CBG}_c$) at hour $t$, see Figure 8 (a). If all elements are known, the hourly outbound for each CBG ($Out_c^t$) is the marginal summation along the $Destination$ dimension, i.e., to collapse the 3D matrix to 2D ($Time$-$Origin$, $\mathbb{R}^{\tau \times n}$, see the blue side in Figure 8 (b)) by aggregating the element values along the $Destination$ dimension. Similarly, the hourly inbound for each CBG ($In_c^t$) is the collapsed 2D matrix of $Time$-$Destination$ along the $Origin$ dimension ($\mathbb{R}^{\tau \times n}$), which is the yellow side of the mobility matrix in Figure 8 (c).

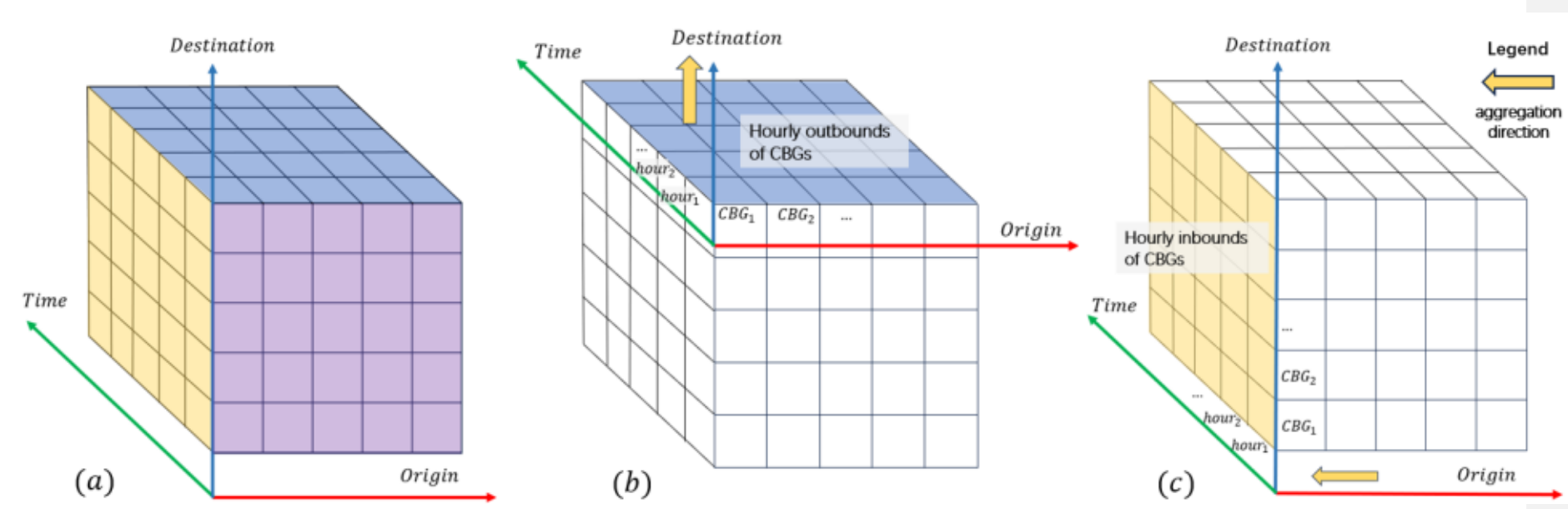

Figure 8 (a): Mobility matrix, whose elements $v_{jc}$ are the visitor counts from $CBG_j$ to $CBG_c$. (b): The aggregations along with the $Destination$ dimension are the hourly outbounds of each CBG, i.e., $Out_c^t$, or the blue side of this sub-figure. (c): The aggregations along with the $Origin$ dimension are the hourly inbound of each CBG, i.e., $In_c^t$, or the yellow side in this sub-figure.

However, the Neighborhood Patterns dataset does not provide the element (i.e., visitor count) for such a 3D mobility matrix; instead, it contains: 1) the hourly stop (not visitor) counts of the destination CBGs, i.e., the marginal summation (1-dimension, $\mathbb{R}^{\tau}$) along the $Origin$ and then $Destination$ dimension; 2) the origin CBGs of the stopped device in destination CBGs in a month, i.e., the collapsed 2D matrix along $Time$ ($Destination$-$Origin$, $\mathbb{R}^{n \times n}$), which is the purple side of the cube in Figure 8 (a). Note that for simplicity in this section, we use the idea of mobility matrix in Figure 8 for visitor count, stop count, and device count, rather than explicitly sketch three different matrices and annotations, respectively.

Therefore, the question of estimating $Out_c^t$ can be formatted as follows: using the hourly CBG stop counts ($\mathbb{R}^{\tau \times n}$) and monthly device origin CBGs ($\mathbb{R}^{n \times n}$), to obtain the marginal 2D summations of the visitor mobility matrix along the $Destination$ dimensions ($\mathbb{R}^{\tau \times n}$), i.e., the blue side in Figure 8 (b). A rigid and straightforward solution is to restore the matrix elements (denoted as $v_{jc}$) and then conduct the marginal summation, after converting the stop counts and device count into visitor counts ($v_{jc}$). One straightforward method of obtaining $v_{jc}$ is to split the hourly stopped devices into origin CBG according to the monthly device home CBG distribution (again, we need to assume that the hourly device home distribution is identical to the monthly device home, see Section 5.2.1); and then, aggregate the $v_{jc}$ into along with the $Destination$ dimension.

However, technically, maintaining this large mobility matrix of $3.5 \times 10^{13}$elements ($\approx 720\ hour \times 220{,}000\ CBG \times 220{,}000\ CBG$) is challenging. Thus, to avoid using intensive computational resources for outbound estimation, we applied the iterative proportional fitting (IPF) algorithm (Lomax and Norman 2016), which estimates matrix element values based on the matrix's marginal summations. We created a 2D hourly outbound matrix ($\mathbb{R}^{\tau \times n}$) with hours as rows and origin CBGs as columns (the blue side in Figure 8). The marginal summations ($\mathbb{R}^{\tau}$) of the hour rows are the sum of all CBG outbounds of each hour, i.e., the summation of the hourly inbound of all destination CBGs, since the total outbound equals the total inbound at each hour. The hourly inbound is estimated using the approach in Section 5.2.1. The marginal summations of origin CBG columns are the sum of all hourly CBG outbounds of each CBG, i.e., the outbound of this calendar month. The IPF was conducted using a Python package named $ipfn$ (Forthommme 2021).

### 5.3 Evaluation

Evaluating the hourly population is challenging since we found rare publicly accessible observations or ground truths. In this study, we utilize the LandScan HD US gridded population dataset, which offers a resolution of approximately 100 meters, as the most suitable reference (Weber et al., 2022). Although not a perfect substitute for ground truth due to its static nature and lack of dynamic hourly changes, LandScan provides a robust framework for comparative analysis.

We used LandScan's daytime and nighttime population, respectively, to compare with our estimated noon and midnight population. Specifically, we compared the weekday average noon population of each month with the LandScan daytime population at the Census blockgroup level; the metric is the relative difference see Equation (3), and the final metric is weighted by LandScan population. We call this metric a difference rather than an error since LandScan is not ground truth, and it is not a dynamic hourly population estimation. Similarly, we compared the midnight estimated population with the LandScan nighttime population.

$$diff_c^{rel} = \frac{\overline{Popu_c^{noon}} - Popu_c^{LandScan}}{Popu_c^{LandScan}} \quad (3)$$

Where**:**

$diff_c^{rel}$: relative difference of the Census block group $c$.

$\overline{\text{Popu}_c^{noon}}$: average weekday noon (12: 00 PM, local time) population of the Census block group $c$ in a calendar month.

$\text{Popu}_c^{LandScan}$: daytime population of the Census block group $c$ in LandScan gridded population data (the latest release is 2021 when conducting this study).

### 5.4 Adjusting coefficient (*k*)

In our study, determining an appropriate value for adjusting coefficient $k$ was crucial to align the estimated Manhattan dynamic population with previous studies, i.e., Moss and Qing (2012)'s work. Moss and Qing (2012)'s work and LandScan 2021 gridded population (Weber et al. 2022).

We aimed to ensure that the weekday daytime peak hour population estimates fall into the range of these established studies. Using a trial-and-error approach, we set the adjusting coefficient $k$ as 4, meaning that the device sampling rate needs to be enlarged by 4 times to roughly include the various uncertainties, such as the unrecorded trips and stops of the panel devices. Figure 9 (a) shows the 2022 hourly population of Manhattan when $k = 4$. Although such a universal and stationary $k$ value cannot fully capture the uncertainties across different hours or cities, we use it in this study pending the integration of further validation data and ground truth observations.

Manhanttan's Census block group level hourly population of a zoomed-in day, Wednesday, January 19, 2022, is mapped in Figure 9 (e). On that day, the hourly population exceeded those recorded by the ACS 2019 population, aligning with the trends reported by Moss and Qing (2012). Most population peaks fall into the range near population peak between the LandScan Daytime and Moss and Qing (2012)'s estimates.

A significant observation was made regarding Central Park, where both ACS and LandScan (Figure 9 (b) and (c) allot no population. Contrary to these two sources, our mobility-based dynamic population identified a consistent visitor hotspot in Central Park, which is more reasonable. Central Park attracts about 42 million visitors annually (Alex Van Buren 2016; Greensward Group 2023; Central Park Conservancy 2022), or 115,000 daily; our estimation is about 300,000 by adding all 24 hourly population a day, while the peak population is mostly about 35,000. It is quite possible that a single visitor who stayed more than 2 hours in the park will be counted in each hour in this research, and the car drivers transversing this CBG through motor roads might also be counted, although they are not park visitors. Therefore, considering that our estimation of 300,000 includes the park visitors and the passengers driving through the CBG, the proposal method is reasonable compared with the park visitor statistics, although there are no straightforward comparing manners between the people count in the statistics and people-visit count in our study.

A significant observation was made regarding Central Park, where both ACS and LandScan allot no population. Contrary to these sources, our mobility-based dynamic population identified a consistent visitor presence in Central Park. Central Park attracts about 42 million visitors annually (Alex Van Buren 2016; Greensward Group 2023; Central Park Conservancy 2022), or 115,000 daily; our estimation is about 300,000 by adding all 24 hourly population a day, while the peak population is mostly about 35,000. Considering such visitors include the passengers driving through the transverse roads across the CBG, the proposal methods and adjusting coefficient $k$ is reasonable compared with the park visitor statistics.

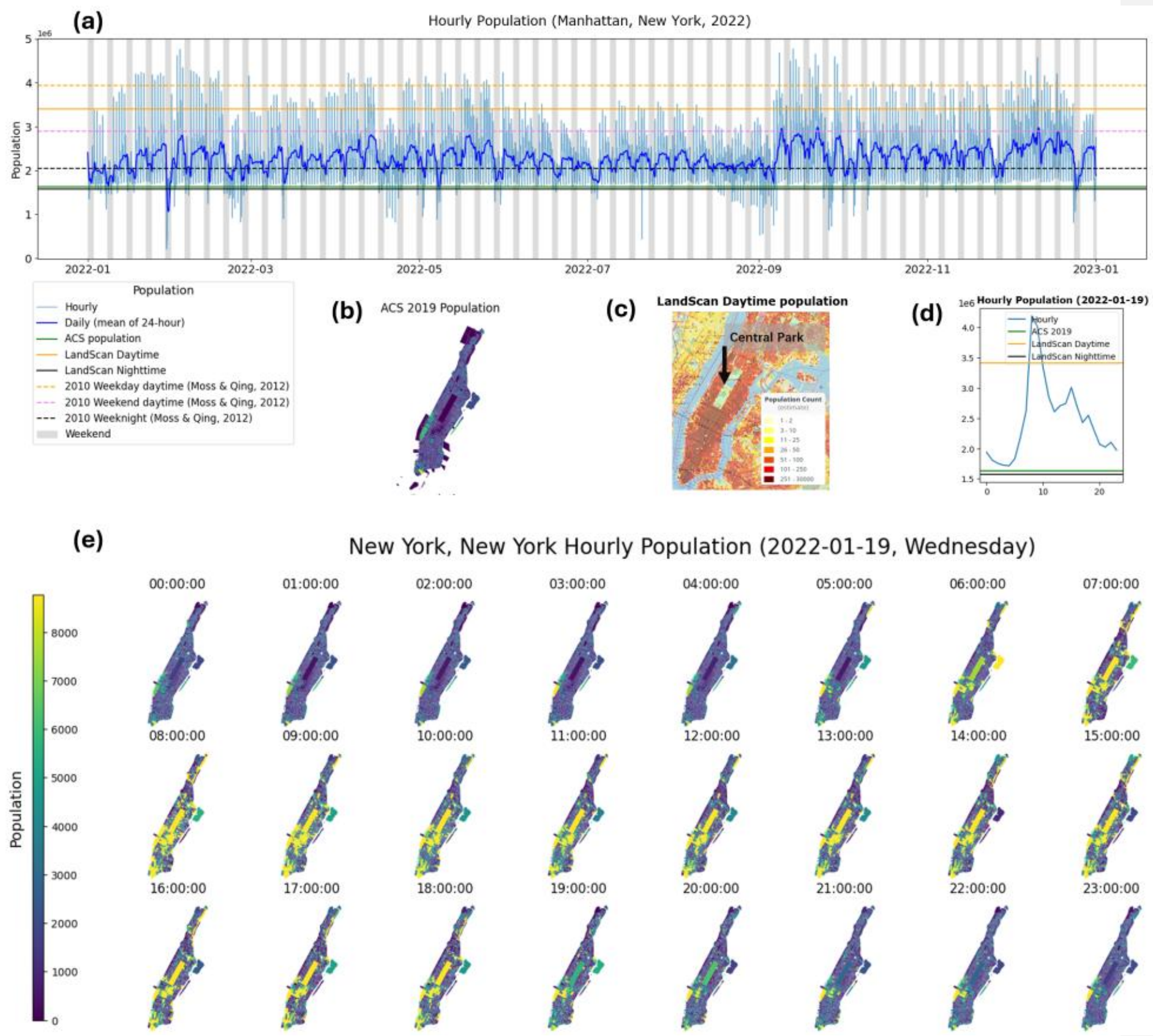

Figure 9 (a): Dynamic population of Manhattan, New York, 2022. (b): Central Park has no residents in the ACS 2019 data. (c) The 2021 LandScan daytime population (landscan.ornl.gov); similar to ACS 2019, the Central Park has no population in daytime, which is unrealistic. (d): The hourly population plot of Manhattan on Wednesday, January 19, 2022. (e): The population dynamics on a weekday for Census block groups in Manhattan, New York. The famous Central Park attracted more than 10000 - 30000 visitors at peak hours.

Note that the adjusting coefficient $k$ is a rough estimation, it may produce a "negative" population in some CBGs (about 10% when $k = 4$), because the device sampling rate, inflated by $k$, overly extracts residents from their original locations, assigning an unrealistic number of individuals to other CBGs. A smaller $k$ value, such as 3, can reduce the CBGs with "negative" population. We eliminated these negative populations by setting a minimum population as 10% of the ACS population; meanwhile, the associated inbounds to destination CBGs were decreased accordingly to keep the balance of the total hourly population, which should equal the ACS population. This balancing is an iterative process and time-consuming; we did this only one time and have 5% CBGs with a negative population. In these instances, we directly adjusted them to the 10% ACS population to simplify the study.

We tried to find an appropriate value for the adjusting coefficient $k$ to let the estimated Manhattan dynamic population be consistent with previous studies, i.e., LandScan (Weber et al. 2022) and Moss and Qing (2012)'s work. We considered more the weekday daytime peak population and tried to make it fall into the range of the mentioned two studies. This Figure shows the 2022 hourly population of Manhattan when $k = 4$. Although such a single stationary value cannot cover the uncertainties of all hours for all cities across the country, we use it in this study before introducing further validation data and ground truth observations.

### 5.5 Evaluation

Using the evaluation method in Section 5, the difference between the mobility-based population and the LandScan population is 67.8% during the daytime and 42.7% for nighttime in 2022 (Table 1). Note that such variations reflect the different aspects of the population, rather than error; for example, the relative difference between the LandScan nighttime and ACS 2019 is 20.6%. The dynamic population patterns revealed by our estimation are reasonable (see Section 2).

Table 1 Difference between mobility-base population and LandScan population

| Month | 1 | 2 | 3 | 4 | 5 | 6 | 7 | 8 | 9 | 10 | 11 | 12 |
|---|---|---|---|---|---|---|---|---|---|---|---|---|
| Noon/daytime (%) | 68.0 | 69.6 | 59.9 | 70.4 | 70.3 | 62.3 | 70.1 | 64.7 | 73.2 | 63.4 | 63.8 | 78.3 |
| Midnight/nighttime (%) | 46.2 | 43.6 | 39.0 | 41.7 | 40.9 | 39.5 | 44.3 | 41.8 | 45.0 | 40.4 | 43.1 | 46.4 |

**Data and code availability statement**

Advan Neighborhoods Patterns can be subscribed from Dewey Inc. (deweydata.io). Census data can be downloaded from U.S. Census Bureau (census.gov).

Data processing code is available on https://github.com/gladcolor/hourly_population. (Code will be uploaded after the paper is accepted or upon request).